\documentclass{article}[12pt]
\usepackage{amsmath,amssymb,amsfonts}
\usepackage{graphics, subfigure,color} 

\usepackage{latexsym}
\usepackage[dvips]{graphicx}
\usepackage{epsfig}
\usepackage{braket}

\usepackage{hyperref}

\numberwithin{equation}{section}

\newcommand{\cZ}{{\cal{Z}}}

\newcommand{\cH}{{\cal H}}

\newcommand{\bee}{\begin{equation}}
\newcommand{\ee}{\end{equation}}
\newcommand{\beq}{\begin{eqnarray}}
\newcommand{\eeq}{\end{eqnarray}}
\newcommand{\bqa}{\begin{eqnarray}}
\newcommand{\eqa}{\end{eqnarray}}
\newcommand{\bea}{\begin{eqnarray}}
\newcommand{\eea}{\end{eqnarray}}
\newcommand{\beann}{\begin{eqnarray*}}
\newcommand{\eeann}{\end{eqnarray*}}  
  
\newcommand{\tr}{{\rm Tr}}

\makeatletter

\begin{document}

\title{Why are tensor field theories asymptotically free?}

\author{
Vincent Rivasseau\footnote{rivass@th.u-psud.fr, Laboratoire de Physique Th\'eorique, CNRS UMR 8627, Universit\'e Paris Sud, 91405 Orsay Cedex, France
 and Perimeter Institute for Theoretical Physics, 31 Caroline St. N, N2L 2Y5, Waterloo, ON, Canada.}}

\maketitle
\begin{abstract} 
In this pedagogic letter we explain the combinatorics underlying the generic asymptotic freedom of tensor field theories. We focus on simple combinatorial 
models with a $1/p^2$ propagator and quartic interactions and on the comparison between the intermediate field representations 
of the vector, matrix and tensor cases. The transition from asymptotic freedom (tensor case) 
to asymptotic safety (matrix case) is related to the crossing symmetry of the matrix vertex,
whereas in the vector case, the lack of asymptotic freedom (``Landau ghost"), as in the ordinary scalar $\phi^4_4$ case, 
is simply due to the absence of any wave function renormalization at one loop.
\end{abstract}

\medskip\noindent
Pacs Numbers: 04.60.-m (Quantum Gravity), 11.10.-z (Field theory), 11.10.Hi (Renormalization group evolution of parameters).

\section{Introduction}
\label{intro}

The Grosse-Wulkenhaar model \cite{Grosse:2004yu}
is a renormalizable non-commutative field theory of the Moyal type
which can also be considered as an abstract matrix model 
with a propagator breaking unitary invariance.  It is asymptotically safe \cite{Disertori:2006nq}. 

In contrast, asymptotic freedom is a generic feature of renormalizable tensor field theories \cite{BenGeloun:2011rc,Geloun:2013saa}
discovered by J. Ben Geloun and its collaborators \cite{BenGeloun:2012pu,BenGeloun:2012yk}.
This remarkable property makes such tensor field theories attractive candidates for quantizing gravity \cite{Rivasseau:2011hm}. 
Just remember how the discovery of asymptotic freedom in non-Abelian gauge theories was the key to
our current understanding of particle physics. However princeps papers such as \cite{BenGeloun:2012yk} are difficult to read, especially since
models with sixth order interactions are treated, and computations are pushed up to four loops, with
subtle issues related to the stability domain and the role
of the relevant fourth order interactions recently underlined \cite{Carrozza:2014rba}.

The purpose of this note is to avoid such technicalities and to better clarify the mechanism responsible for this asymptotic freedom of tensor models, 
concentrating for pedagogical reasons on the computation of the one-loop $\beta$ function of the simplest 
renormalizable combinatorial models with a $1/p^2$ propagator and quartic interactions.

We shall compare the intermediate field representation of such models in the vector, matrix and tensor case.
Indeed this representation simplifies the combinatorics involved and illuminates
why the sign of the one-loop beta coefficient is positive for vectors, 
zero for matrices and negative for tensors. 


\section{Quartic Models and their Intermediate Field Representation}

Consider a pair of complex conjugate fields $\phi (\theta)$ and $\bar \phi (\bar \theta)$ on the $d$-dimensional torus $T_d = S_1^d$,
hence $\theta = (\theta_1, \cdots , \theta_d) $, $\theta_c\in S_1$. Their Fourier series will be noted as
$\phi_p$ and $\bar \phi_{\bar p}$ where $p =   (p_1, \cdots , p_d),\; \bar p  =  (\bar p_1, \cdots , \bar p_d) \in \mathbb{Z}^d$.
The corresponding Hilbert space is $\cH_d = L^2 (S_1^d) = \ell_2 ( {\mathbb Z}^d) = \otimes_{c=1}^d  \cH_c$, where each
space $\cH_c = L^2 (S_1) = \ell_2 ( {\mathbb Z})$ corresponds to a given component also called \emph{color} in the tensorial context \cite{Gurau:2009tw,Gurau:2011xp}.

We introduce the familiar Laplacian-based normalized Gaussian measure\footnote{This propagator is the familiar one of ordinary local scalar Bosonic field theory,
but is also natural in the more abstract combinatorial context, see eg \cite{Geloun:2011cy}.} for $d$-dimensional Bosonic fields with periodic boundary conditions
\bee
d\mu_C(\phi, \bar \phi) = \left(\prod_{p, \bar p \in \mathbb{Z}^d} \frac{d\phi_p d\bar \phi_{\bar p}}{2i\pi} \right) \mathrm{Det}( C )^{-1} \ 
e^{-\sum_{p, \bar p} \phi_p C^{-1}_{p\bar p}\bar \phi_{\bar p}}
\ee
where the covariance $C$ is, up to a field strength renormalization, the inverse of the Laplacian on $S_1^d$ plus a mass term
\bee
C_{p,\bar p}=\frac{1}{Z} \frac{\delta_{p,\bar p}}{p^2+ m^2}.
\ee
Here $p^2 = \sum_{c=1}^d p_c^2 $, $m^2$ is the square of the bare mass, and $Z$ is the so-called
wave-function renormalization, which can be absorbed into a $(\phi, \bar \phi) \to (Z^{-1/2} \phi, Z^{-1/2} \bar \phi)$ field strength renormalization. 
If we restrict the indices $p$, which should be thought as``momenta", to lie in $[-N, N]^d$ rather than in $\mathbb{Z}^d$ we have proper (finite dimensional) fields.
We can consider $N$ as the ultraviolet cutoff, and we are interested in performing the ultraviolet limit $N \to \infty$.

The generating function for the moments of the model is
\bee
\cZ(g,J, \bar J)=  \frac{1}{\cZ}\int e^{\bar J \cdot \phi + J \cdot \bar \phi }   e^{-\frac{g}{2}V (\phi, \bar \phi) } d\mu_C(\phi, \bar \phi) , \label{action}
\ee
where $\cZ= \cZ(g,J, \bar J)\vert_{J =\bar J =0}$ is the normalization, $g$ is the coupling constant\footnote{The factor $1/2$ in front of $g$ takes care of the ${\mathbb{Z}}_2$ symmetry of 
the quartic vertices considered below, hence simplifies their associated intermediate field
representation.}, and the sources $J$ and $\bar J$
are dual respectively to $\bar \phi$ and $\phi$. The generating function for the connected moments is $W = \log  \cZ(g,J, \bar J)$.

The quartic vector, matrix and tensor \emph{combinatorial} field theories correspond to
choices of the quartic interaction $V (\phi, \bar \phi)$ which simply
differ in the way the momenta indices of the four fields branch at  $V$. They correspond to different symmetry groups,
and are not \emph{local} in the usual sense on $S_1^d$. For instance the tensorial case corresponds to interactions invariant under
$U(2N+1)^{\otimes d}$, hence under \emph{independent} change of basis in $\cH_c$ for each color $c$ \cite{Gurau:2011kk,Bonzom:2012hw}.
Such combinatorial interactions are interesting in the context of quantization of gravity, since the
corresponding Feynman graphs are dual to $d$-dimensional (colored) triangulations \cite{Gurau:2010nd} pondered by a discretized form of the Einstein-Hilbert action \cite{ambjorn}.

The \emph{vector} interaction is the square of the quadratic (mass) term, hence is factorized\footnote{There
are no connected polynomial invariants for vectors beyond the scalar product.} (not connected). It writes
\bee
V_V =  <\bar \phi , \phi>^2 =\sum_{p, q} \left( \phi_p \bar \phi_{p} \right)\left( \phi_q \bar \phi_{q} \right), \label{quartvect}
\ee
and it is just renormalizable for $d=4$.

The \emph{matrix} interaction makes sense only for $d=2r$ even and is obtained by splitting the initial index as a pair $(p,q)$ with $p = (p_1 , \cdots, p_r) $, $q= (q_1 , \cdots, q_r) $, hence
splitting the space $\cH_d = \cH_r \otimes \cH_r$. The field $\phi$ is then interpreted as the matrix $\phi_{p q}$, with conjugate matrix 
$\phi^\star = ^t\bar \phi$ and the vertex $V_M$ is an invariant trace
\bee
V_M = \tr \; \phi \, \phi^\star \phi \, \phi^\star  = \sum_{p,q, p',q'} \phi_{pq} \bar \phi_{p'q} \phi_{p'q'} \bar \phi_{pq'}. \label{quartmatr}
\ee
It it is just renormalizable for $r=4$, hence $d=8$.

Finally the simplest \emph{tensor} interaction $V_T$ is the color-symmetric sum of \emph{melonic} \cite{Bonzom:2011zz,Gurau:2011xp} quartic interactions \cite{Gurau:2013pca,Delepouve:2014bma}
\bea
&&V_T = \sum_c V_c,  \quad  V_c(\phi, \bar \phi) = \tr_c ( \tr_{\hat c}  \phi \bar \phi )^2  \label{quarttens} \\
&&=
\sum_{p,\bar p, q,\bar q} \biggl[ \phi_p\bar \phi_{\bar p}  \prod_{c'\neq c}\delta_{p_{c'} \bar p_{c'}} \biggr] \delta_{p_c \bar q_c} \delta_{q_c \bar p_c} 
\biggl[ \phi_q\bar \phi_{\bar q} \prod_{c'\neq c}\delta_{q_{c'} \bar q_{c'}}  \biggr] , \nonumber 
\eea
where $\tr_{\hat c}  \phi \bar \phi$ means partial trace in $\cH_d = \otimes_{c=1}^d  \cH_c$ over all colors except $c$, and $\tr_c$ means trace over color $c$.
The corresponding model is just renormalizable for $d=5$  \cite{Samary:2014oya,Avohou}, which we now assume in this case.

These three different combinatorial models have just renormalizable power counting, like the ordinary ``scalar" $\phi^4_4$. 
But the class of divergent graphs is more restricted in the combinatorial case. 

Remark that in all cases the interactions $V$ are positive for $g >0$. Hence the models are stable 
for this sign of the coupling constant, which we now assume. Their perturbative series are certainly at least Borel summable at finite cutoff $N$,
and in some super-renormalizable cases have been shown to remain Borel summable 
in the $N \to \infty$ limit \cite{Gurau:2013oqa,Delepouve:2014hfa}.

We shall now compare the one-loop beta function of these models. For this we first pass to the intermediate field 
representation\footnote{This representation is called the Hubbard-Stratonovic 
representation in condensed matter.}. Indeed
quartic models, and especially quartic vector and tensor models,  simplify in this representation: their dominant graphs as $N \to \infty$ simply become \emph{trees}\footnote{More precisely, ordinary trees
in the vector case and colored trees in the tensorial case \cite{Gurau:2013pca}.}. 
The guiding principle is to introduce an intermediate field $\sigma$ to split the quartic vertex in two halves, as pictured in fig.~1, through the simple integral representation
\bee   e^{-\frac{g}{2}V (\phi, \bar \phi) } =   \int d\nu(\sigma)  e^{i \sqrt g  \bar \phi\phi \cdot \sigma} \label{interm1} .
\ee
In this formula $d\nu$ is a Gaussian measure with covariance 1 on the intermediate field $\sigma$. The three cases \eqref{quartvect}-\eqref{quarttens} lead
to $\sigma$ fields of different nature and to different combinatorial `rules for the dot" in \eqref{interm1}. 
In the vector case the $\sigma$ field is a \emph{scalar}, reflecting the already factorized nature of  \eqref{quartvect}. In the matrix and tensor case 
\eqref{quartmatr}-\eqref{quarttens} $\sigma$ is a \emph{matrix}. More precisely,  in the matrix case \eqref{quartmatr}, it is a single matrix with its two arguments in ${\mathbb Z}^4$. In
the tensor case \eqref{quarttens}, it is the sum of five different 
``colored matrices" $\sigma_c$ with their two arguments in ${\mathbb Z}$, one for each color $c$, and should be properly written as \cite{Gurau:2013pca}
\bea {\sigma }&=& \sum_c   \sigma_c \otimes \mathbb{I}_{\hat c} =
\sigma_1\otimes\mathbb{I}_2\otimes\mathbb{I}_3\otimes\mathbb{I}_4\otimes\mathbb{I}_5
\cdots + \mathbb{I}_1\otimes\mathbb{I}_2\otimes\mathbb{I}_3\otimes\mathbb{I}_4\otimes \sigma_5.
\eea

\begin{figure}[!t]\label{cutvertex}
  \begin{center}
  {\includegraphics[width=0.6\textwidth]{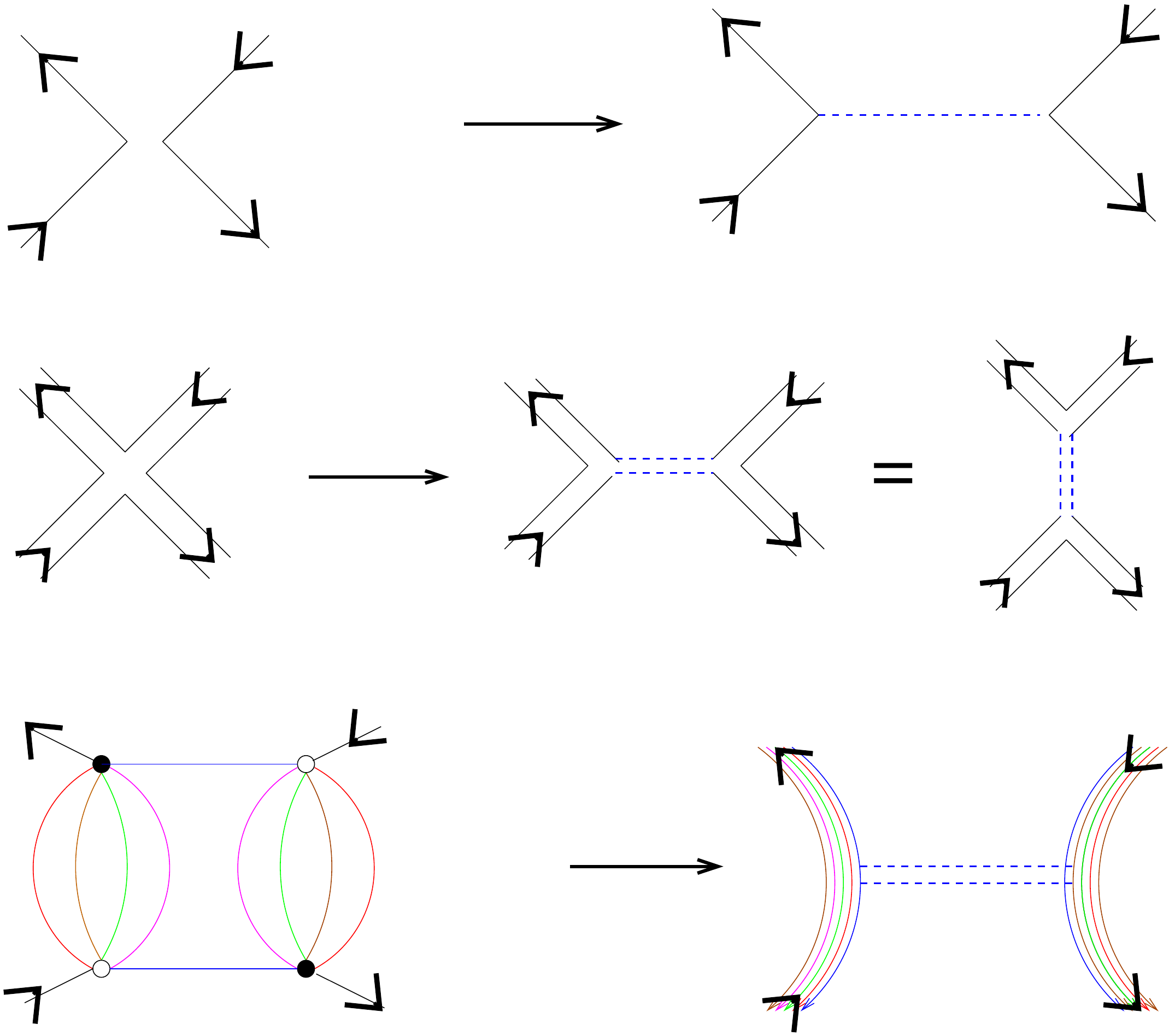}}   
   \end{center}
  \caption{The vertex is cut in two by the intermediate field representation. From top to bottom: the vector, matrix and tensor case.
  Incoming and outgoing arrows distinguish $\phi$ and $\bar \phi$.}

\end{figure} 

The advantage of this representation is that the $\bar\phi$ and $\phi$ functional integral becomes quadratic, hence can be
performed explicitly, yielding 
\begin{eqnarray}
\cZ(g,J, \bar J)&=&  \frac{1}{\cZ} \int d\nu(\sigma)   \int d\mu_{C}(\phi, \bar \phi)    e^{i \sqrt g  \bar \phi\phi \cdot \sigma}
e^{\bar J \cdot \phi + J \cdot \bar \phi }  \nonumber  \\
&=&   \frac{1}{\cZ}  \int  d\nu(\sigma) e^{< \bar J ,  C^{1/2}  R (\sigma) C^{1/2}  J > - \tr \log \left[\mathbb{I}-i  \sqrt g  C^{1/2} {\sigma } C^{1/2} \right]  }  ,
 \label{log2}
\end{eqnarray}
where $R$ is the symmetric \emph{resolvent} operator
\bee 
R  (\sigma)  \equiv \frac{1}{  \mathbb{I} -i  \sqrt g  C^{1/2}  \sigma  C^{1/2} } .
\ee
Writing symmetrized expressions with $C^{1/2}$'s is a bit heavier but shows the Hermitian nature of $C^{1/2}\sigma C^{1/2} $. It is essential to ensure
the non-perturbative existence of the resolvent and logarithm for $g$ in a \emph{cardioid domain} of the complex plane, see e.g. \cite{Gurau:2013pca,Delepouve:2014bma,Delepouve:2014hfa}.

\section{One Loop $\beta$ Function}

In all cases the one-loop beta function boils down to the same computation, up to subtle differences 
of purely combinatorial nature. Let us call $\Gamma_{2p}$ the $2p$-point vertex function, hence
the sum of one-particle irreducible amputated Feynman amplitudes with $2p$ external legs. 
The renormalized BPHZ prescriptions are defined by
momentum space subtractions at zero momentum\footnote{In fact we shall restrict these subtractions to \emph{divergent} graphs (e.g. planar in the matrix case, melonic in the tensor case), 
the difference being a finite renormalization.}
\bee
\frac{g_r}{2} = - \Gamma_{4}(0),\quad Z-1 = [ \frac{\partial}{\partial p^2}  \Gamma_2 ] (0), \label{gammas}
\ee
where $g_r$ is the renormalized coupling. Performing the field strength renormalization we can rescale to 1 the wave function renormalization at high ultraviolet cutoff at the cost
of using a \emph{rescaled} bare coupling  $g'_b = Z^{-2} g_b$. The one-loop $\beta_2$ coefficient shows how this rescaled bare coupling evolves at fixed $g_r$ when $N \to \infty$.
It writes
\bee  g'_b =  g_r [1 + \beta_2 g_r ( \log N + {\rm finite}) + O(g_r^2) ], \label{beta}
\ee
where $N$ is the ultraviolet cutoff, and ``finite" means a function which is bounded as
$N \to \infty$.
As well-known $\beta_2 > 0$ corresponds to a coupling constant which flows out of the perturbative regime in the ultraviolet (ÒLandau ghostÓ). $\beta_2 < 0$ corresponds to the nice physical
situation of asymptotic freedom: the (rescaled) bare coupling flows to zero as $N \to \infty $, hence the
(rescaled) theory tends towards a Gaussian (free) quantum free theory in the ultraviolet regime. $\beta_2 =0$
indicates the possibility of asymptotic safety (non trivial ultra-violet fixed point close to the Gaussian one), but 
is inconclusive as the analysis of the renormalization group flow must be pushed further.

It is easier to compute the bare perturbation theory, as it does not involve any subtraction.
Starting from \eqref{gammas}, we find that $\Gamma_{4}$ and $Z-1$ always involve the \emph{same} logarithmically divergent sum, 
namely (using e.g. the standard parametric representation \cite{Avohou})
\bee
\sum_{q\in[-N,N]^4}\frac{1}{(q^2+m_r^2)^2} = 2 \pi^2 \log N + {\rm finite}  , \label{beta3}
\ee
where $m_r^2= Zm^2 -\Gamma_{2} (0)$ is the renormalized mass. 
However this sum arises with various combinatoric coefficients. More precisely
\bee
\Gamma_{4} (0)  = -\frac{g_b}{2} [1 - a g_b \sum_{q\in[-N,N]^4}\frac{1}{(q^2+m_r^2)^2}+O(g_b^2)] , \label{beta1} 
\ee
\bea
Z&=& 1 + \frac{\partial \Gamma_{2}}{\partial p^2} \Big|_{p=0} \;=\;  1 + b g_b \sum_{q\in[-N,N]^4}\frac{1}{(q^2+m_r^2)^2}+O(g_b^2),  \label{beta2}
\eea
where $a$ and $b$ are combinatoric coefficients that depend on the particular case (vector, matrix or tensor) considered.

Since $g'_b = Z^{-2} g_b$, multiplying \eqref{beta} by $Z^2$ and taking into account \eqref{gammas}-\eqref{beta2}, which imply $g_r = g_b + O(g_b^2)$ and $Z= 1 + O(g_b)$, we find
\bee  Z^2\Gamma_4 (0) = -\frac{g_b}{2} [1 - \beta_2 g_b ( \log N + {\rm finite}) + O(g_b^2) ],
\ee
hence in all cases we find that
\bee  \beta_2 =  (a-2b)  2 \pi^2 ,
\ee
and we are left with the simple problem of computing the coefficients $a$ and $b$ of the one loop leading diagrams 
for $\Gamma_4$ and $Z$.

\begin{figure}[!t]
  \begin{center}
  {\includegraphics[width=0.6\textwidth]{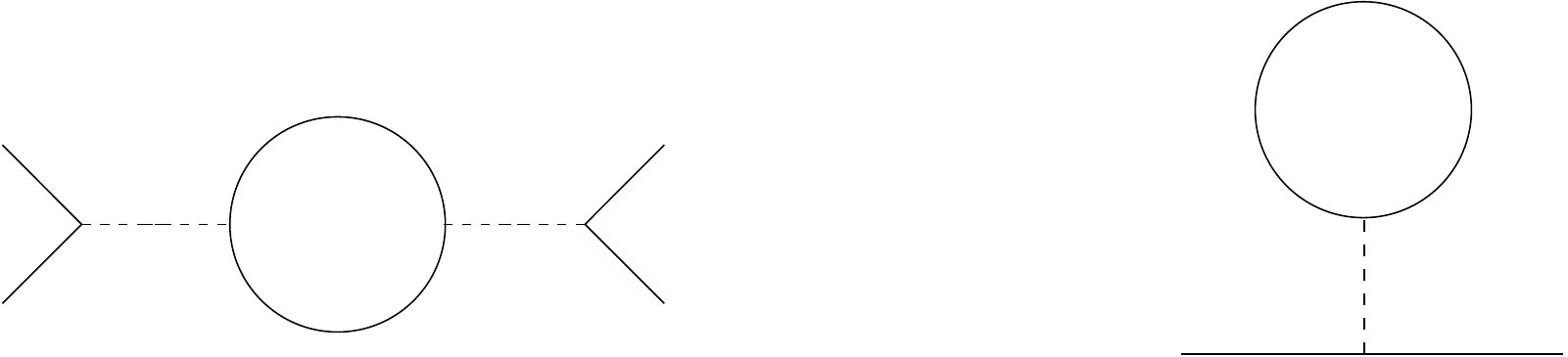}}
   \end{center}
  \caption{The (single) one loop melonic graph in the tensor case for $\Gamma_4$ and $\Gamma_2$ are trees for the intermediate field (dashed) lines.}
  \label{onelooptensor}
\end{figure}

\begin{itemize}

\item
In the vector case, $a=1$. Indeed in the parametric representation the only divergent graph is the one on the left of fig.~2. 
Resolvents (derivatives with respect to $J$ and $\bar J$), also represented as ciliated vertices in \cite{Gurau:2013pca},  
come up with a factor 1 and no symmetry factorial, whether terms from the $\tr \log$ expansion, also called \emph{loop vertices} \cite{Rivasseau:2007fr}
and pictured as unciliated vertices in \cite{Gurau:2013pca} come up with a factor $1/n$ for a $ \tr ( i \sqrt g C^{1/2} {\vec \sigma }C^{1/2})^n $ (because of the Taylor series of the logarithm), 
plus a symmetry factor $1/k!$ if there are $k$ of them (this factor comes from expansion of the exponential). 
The combinatoric weight for the tree graph at order $ g^2$ for $\Gamma_4$ is therefore $1$, which decomposes into 
a $1/2$ for the single loop vertex ($n=2, k=1$) times a 2 for the two Wick contractions.

In this vector case $b=0$ since the one loop tadpole on the right of fig.~2, the only contributing graph at order $g$, does not have any 
external momentum dependence. Hence 
\bee [ \frac{\partial}{\partial p^2}  \Gamma_2 ] (0) = O(g^2)\quad  => \quad b=0.
\ee
Hence $\beta_2 = 2 \pi^2  $ and the theory has no UV fixed point, at least in this approximation.

\item In the tensor case, $a=1$ for the same reasons than in the vector case. Indeed for any of the five melonic interactions there is a single
divergent graph of the corresponding color of the type pictured on the left of fig.~2. 

But now we also have $b=1$. Indeed remark first that $b>0$ because 
two minus signs compensate, one in front of $g$ in \eqref{action} and the other coming from the mass subtraction, since 
\bea
&&\biggl[\frac{1}{p_c^2}   \bigl( \frac{1}{q^2 + p_c^2  + m_r^2}  - \frac{1}{q^2  + m_r^2}  \bigr)  \biggr]_{p_c = 0} \nonumber \\
&=& - \frac{1}{ ( q^2 + m_r^2 )^2} .
\eea

The combinatorics is then 1 because there is a single loop vertex with $n=1, k=1$ and a single Wick contraction to branch it on the external resolvent
as shown for the graph on the right of fig.~2.
Summing over the colors $c$ of this contraction simply reconstructs $\sum_c p_c^2 =p^2$. In conclusion
$\beta_2 = -2 \pi^2 $ and the theory is asymptotically free,  in agreement with  
\cite{BenGeloun:2012pu,Samary:2014oya,Benedetti:2014qsa}. Wave function, or field strength renormalization won over 
coupling constant renormalization because of the square power in $Z^2$.

\begin{figure}[!t]
  \begin{center}
  {\includegraphics[width=0.65\textwidth]{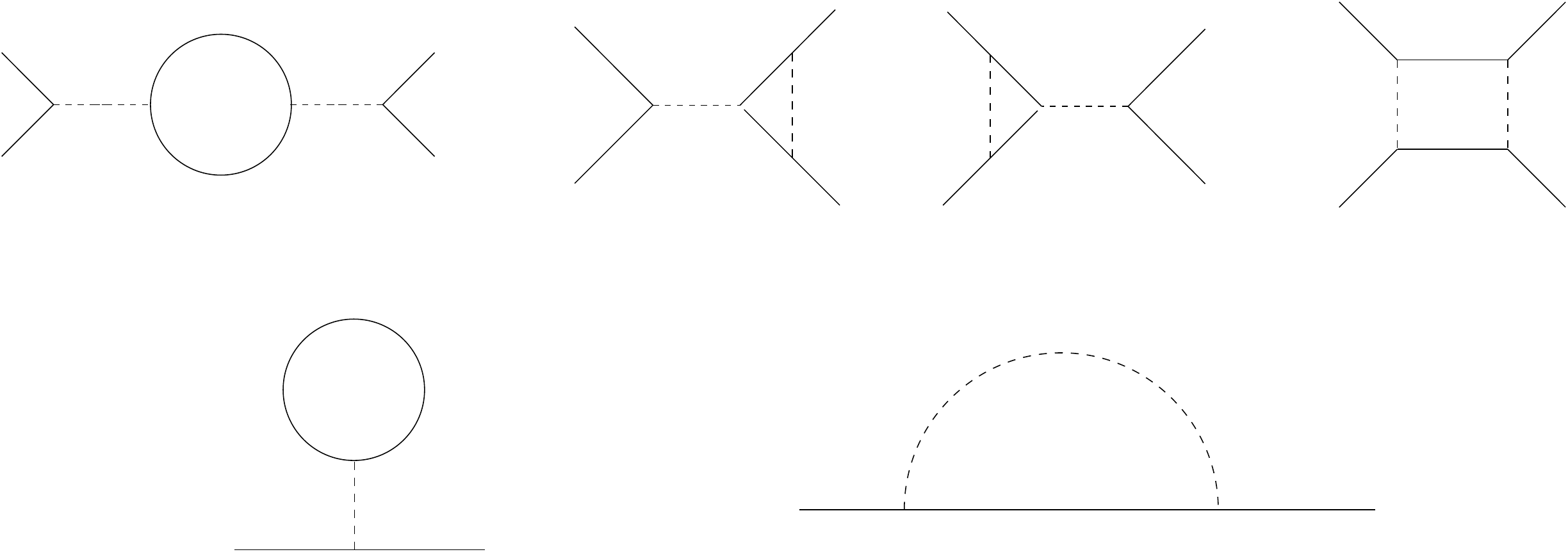}}
   \end{center}
  \caption{The dominant one loop graphs in the matrix case for $\Gamma_4$ and $\Gamma_2$ are planar in the intermediate field representation.}
  \label{oneloopmatrix}
\end{figure}

\item In the matrix case, the vertex crossing symmetry means more terms diverge logarithmically than in the vector 
and tensor cases, namely those corresponding to planar \emph{maps}\footnote{As by now well known, intermediate field graphs are really what mathematicians 
nowadays call combinatorial maps.} in the intermediate representation.
The crossing symmetry is a ${\mathbb Z}_2$ symmetry, but it acts differently on $\Gamma_4$ and $\Gamma_2$.
Since the one loop graph for $\Gamma_4$ has \emph{two} vertices, hence two (dotted) $\sigma$ propagators, the crossing symmetry acts twice independently 
and generates an orbit of four planar maps, represented in the top part of fig.~3. In contrast the crossing symmetry acts only once on
the orbit of the $\Gamma_2$ term, generating only the two planar maps pictured in the bottom part of fig.~3. Hence
$a=4$ and $b=2$ which leads to $\beta_2 = 0$!  This ``miracle" persists at all orders: in fact the logarithmically divergent part of 
$ Z^2 \Gamma_4 (0)$ is exactly 0 at all orders in $g$, as can be shown through combining a Ward identity with the Schwinger Dyson equations of the theory 
\cite{Disertori:2006nq}. The corresponding theory is asymptotically \emph{safe}.

\end{itemize}

In conclusion the asymptotic freedom of quartic melonic renormalizable tensor field theories can be traced back to the same combinatorial weight 1 for the melonic 
terms (rooted trees in the IF representation) contributing to $\Gamma_4 (0) $ and $Z$. It leads to $Z^2$ winning over  $\Gamma_4 (0) $. In contrast
asymptotic safety for matrix models comes from the crucial crossing symmetry of the vertex. This symmetry 
boosts the contribution of $Z$ by a factor 2, but the contribution of $\Gamma_4 (0) $ by a factor 4, since it acts independently 
on its two vertices. This restores perfect equilibrium between  $\Gamma_4 (0) $ and $Z^2$.

\section{Some Further Remarks}

The reader could ask where in this picture sits the ordinary scalar (complex) $\phi^4_4$ theory. It corresponds to
a vertex which is local in the direct $\theta $ representation, hence momentum conserving in the $n$ indices. 
This vertex in the language of the previous section is therefore, at $d=4$, hence for $p , \bar p, q, \bar q \in {\mathbb Z}^4$
\bee
V_S  =\sum_{p,\bar p, q,\bar q}   \phi_p \bar \phi_{\bar p}  \phi_q \bar \phi_{\bar q} \delta (p - \bar p + q - \bar q ).
\ee
The symmetry of the vertex no longer distinguishes planar from non planar Wick contractions. Moreover $b=0$, like in the vector case, because the tadpole in this theory
is purely local. Again there is no way to have asymptotic freedom or even safety for a stable $g>0$ coupling, because there is no one-loop wave function renormalization.

One could also consider real rather than complex models. Again here there are some changes, for instance the crossing symmetry
of the scalar vertex has a symmetry of order three, since there are three ways to divide the four equivalent real fields into two pairs.
But the conclusions do not change: real scalar and real vector theories are neither asymptotically free nor safe
since there is still no one-loop wave function renormalization, whether renormalizable real matrix theories 
remain  asymptotically safe (the combinatorics being the same than for the real matrix Groose-Wulkenhaar model). 
We expect also real tensor theories to remain asymptotically free, since the quartic vertex pairing still has only one melonic channel.

These features also do not depend as much as one could think of the particular form of the propagator. In particular if we replace ${\mathbb Z}$ by ${\mathbb N}$
and considers linear type inverse propagators $\frac{1}{n + m^2}$ the conclusions remain similar. 
The Grosse-Wulkenhaar is a matrix model with such an inverse-linear propagator and is asymptotically safe at all orders \cite{Disertori:2006nq},
and the quartic melonic tensor model with an inverse-linear propagator is asymptotically free in the dimension $d=3$ where it is just renormalizable
\cite{BenGeloun:2012pu}.

Finally let us remark that adding Boulatov-type gauge projectors to the tensor propagators, as is natural in tensor
group field theory \cite{Carrozza:2012uv,Carrozza:2013wda,Carrozza,Lahoche:2015ola,Lahoche:2015tqa} \emph{enhances} rather than suppresses their asymptotic safety. In the simplest 
model of this type which is just renormalizable, namely the quartic melonic $d=6$ model with propagator $\delta( \sum_{c=1}^6 p_c) (p^2 + m^2)^{-1}$ \cite{Lahoche:2015ola}
the basic one-loop integral \eqref{beta3} is changed into
\bee
\sum_{q\in[-N,N]^5}\frac{\delta (\sum_{c=1}^5 q_c )}{(q^2+m_r^2)^2} = \frac{2\pi^2}{\sqrt{5}} \log N + {\rm finite}    . \label{beta4}
\ee
The derivative contributing to $Z$ is also enhanced, 
so that the ratio between $b$ and $a$ is now $6/5$ instead of 1\footnote{Beware of a missing factor 2
in the initial version of that computation \cite{Lahoche:2015ola}.}. As
a result $Z^2$ wins even \emph{more} over $\Gamma_4$. In the normalization used in the previous section we have for that model 
\bee  \beta_2 = a -2b = (1-\frac{12}{5})  \frac{2\pi^2}{\sqrt{5}}  = - \frac{14\pi^2}{5\sqrt{5}}  .
\ee

In conclusion the asymptotic freedom of tensor field theories at rank $d\ge 3$
is a stable mathematical fact, relatively independent of the details of the model, which is 
deeply rooted in its combinatorics, different from the one of matrix models. It would be interesting to study whether, like for non-Abelian gauge theories,
this asymptotic freedom survives only to the addition of a bounded number of matter fields. 


\end{document}